# Exploring AI in Steganography and Steganalysis: Trends, Clusters, and Sustainable Development Potential

Aditya Kumar Sahu, Chandan Kumar, Saksham Kumar, Serdar Solak

**Abstract**

Steganography and steganalysis are strongly related subjects of information security. While steganography is viewed as concealing communications, steganalysis is concerned with the detection of concealed data, and if feasible, its recovery. Over the past decade, many powerful and efficient artificial intelligence (AI) - driven techniques have been designed and presented during research into steganography as well as steganalysis. This study presents a scientometric analysis of AI-driven steganography-based data hiding techniques using a thematic modelling approach. A total of 654 articles within the time span of 2017 to 2023 have been considered. Experimental evaluation of the study reveals that 69% of published articles are from Asian countries. The China is on top (TP:312), followed by India (TP-114). The study mainly identifies seven thematic clusters: steganographic image data hiding, deep image steganalysis, neural watermark robustness, linguistic steganography models, speech steganalysis algorithms, covert communication networks, and video steganography techniques. The proposed study also assesses the scope of AI-steganography under the purview of sustainable development goals (SDGs) to present the interdisciplinary reciprocity between them. It has been observed that only 18 of the 654 articles are aligned with one of the SDGs, which shows that limited studies conducted in alignment with SDG goals. SDG9 which is Industry, Innovation, and Infrastructure is leading among 18 SDGs mapped articles. To the top of our insight, this study is the unique one to present a scientometric study on AI-driven steganography-based data hiding techniques. In the context of descriptive statistics, the study breaks down the underlying causes of observed trends, including the influence of DL developments, trends in East Asia and maturity of foundational methods. The work also stresses upon the critical gaps in societal alignment, particularly the SDGs, ultimately working on unveiling the field's global impact on AI security challenges

## 1. Introduction

Since the advent of the information age, the protection of sensitive data from unethical access is of paramount importance. Over the recent years, scientists have developed novel security methods to combat new and advanced threats within conventional InfoSec techniques. Predominantly, almost all security techniques belong to one of the three approaches, such as (1) cryptography, (2) steganography, and (3) watermarking. Despite the fundamental objectives of all three security approaches being identical, the operating philosophy is quite divergent. Cryptography is the process of enciphering and deciphering a message using strong mathematical methods (Yuan et al., 2024.). Concurrently, watermarking is the process of concealing visible or invisible identification marks in any digital object (Nawaz et al., 2024). In the same way, steganography is a stealth communication. In the ancient past, steganography was primarily performed physically (Kombrink et al., 2024). Steganography is an indispensable tool in the field of cybersecurity. It is meant to covertly protect, secure, and communicate confidential information in various digital formats (Zeng et al., 2024). The guiding principle behind steganographic-based covert communication is to obscure data in a carrier multimedia entity, so that one apart from the intended recipient can comprehend the presence of data in the received object. The importance of steganography can be well understood because it provides robustness, enhanced security towards confidential communication, protection of intellectual property, data integrity and authentication, safe information storage, etc (Zhang et al., 2023). Steganography conceals the very existence of any information in a digital object. It could be either image, audio, video, or text. After concealing the information, the digital entity is typically known as a 'stego-object'. The essential feature of a stego-object is to be an indistinguishable object from the carrier object to deceive the opponent.

Over the past three decades, the evolution of steganography has progressed in tandem with steganalysis. In this pursuit of cover and uncover, there is an infinite race between steganography and steganalysis. A steganographer always looks to achieve higher embedding capacity with lower perceptual distortion (Meng et al., 2024). At the same time, during steganalysis, it tries to map the presence of the masked information within the stego-object. Recently, reversible steganography approaches have gained credible attention because of their ability to recover the original carrier object at the recipient end (Zhang et al., 2024). However, owing to the remarkable growth of IoT and the significant advancement in AI, steganography has advanced from the classical method of hiding information to the contemporary approach. In steganography contexts, AI is used to embed confidential data within images. Steganography hides data, while steganalysis reverse-engineers the hidden information. The recently developed AI frameworks can be used to steganalyse cryptic images (Gupta et al., 2021). Steganography is often used for clandestine transmission, maintaining the confidentiality and protection of secret data (Mukherjee et al., 2023). Novel Techniques based on Neural Nets have been created to obscure confidential data within images, each with its own set of pros and cons (Chanchal et al., 2020). LLM models can be used to formulate steganographic models with the power of comprehensive literature analysis (Mukherjee et al., 2023). Integration of AI and steganography has the power to fuse InfoSec and safeguard against harmful attacks (Al-Hussein et al., 2023). As observed, AI plays a pivotal role in improving the security and efficiency of Steganography, by analysis, ideation, review, and guardrail practices for testing.

Cheddad et al (2010) provided an expansive review of steganography-based applications, from barcodes to and digital forensics to forging documents. It also discusses many established techniques, including format exploitation and image frequency domain analysis. Kaur et al. (2022) also conducted a systematic review of image-steganography works, comparing their depths and flaws. They also assess blockchain-based steganography techniques to determine their potential as a superior stego medium. Mandal et al. (2022) studied some recent steganography strategies and explored widely used steganography tools comprehensively. Majeed et al. (2021) reviewed existing works on the development of approaches and algorithms for text steganography only. Zhou et al. (2021) conducted a systematic survey on 3-D mesh steganography and steganalysis. The authors proposed a new taxonomy of steganographic and steganalysis algorithms. They also discussed the history of technical advancements and the current technological level. Dalal and Juneja (2021) conducted a holistic analysis incorporating qualitative and quantitative data from various video steganography approaches. They highlighted the real-life applications of video steganography. Rustad et al. (2023) conducted a review on image steganography. They covered steganography's goal, assessment, method, development, and dataset. This review aims to classify steganography based on goals and assessment, review the use of assessment tools and challenges or issues of steganography development, as well as provide an investigation, critical analysis, and an explicit summary for novice researchers to understand image steganography development. Meng et al. (2023) conducted a review of coverless steganographic algorithms, featuring the noteworthy improvement in coverless image and video steganographic methods.

Literature suggests that different text-based, image-based, video-based, and 3-D mesh steganography reviews have been conducted in the past. However, no researchers have conducted a comprehensive review of AI-steganography focusing on thematic modeling. Therefore, a comprehensive and focused systematic literature review becomes imperative to address and provide a thorough understanding of the research landscape in AI-driven steganography.

### 1.1 The need for a comprehensive scientometric analysis of AI-steganography

These are some key aspects of the scientometric analysis of AI-based steganography.

1. **A comprehensive evaluation of AI-based steganography techniques**: AI-steganography involves various methods, including data-driven learning, deep learning, linguistic computing, convolutional network (CNN), generative adversarial networks (GAN), deep neural networks(DNN), cloud computing, and so on. Past studies on steganography and steganalysis have been denounced for their restricted scope, lack of multi-faceted and diversified perspectives, and need for more objective, technology-facilitated analytical methods. This systematic review will highlight the most influential themes and topics of AI-steganography.

2. **Identification of research gaps:** AI-steganography is rapidly evolving, with new techniques and applications emerging frequently. This systematic review allows researchers to identify areas that require further investigation, encouraging the development of novel approaches and technologies.

Understanding publication and citation trends in AI-steganography is beyond bibliometric accounting. It is a strategic tool for multiple stakeholders in the research landscape. For the research agenda, analysing these trends reveals methodologies that are rarely gaining traction, which approaches are becoming obsolete, and where novel opportunities lie—enabling researchers to position their work at the frontier of the field rather than revisiting saturated topics. From a funding strategy, identifying high-impact clusters, productive institutions, and emerging themes helps funding agencies allocate resources efficiently to areas with the greatest potential for breakthrough discoveries or societal impact. From the view of policymakers and authorities, by summarising the developments of AI-driven steganography, adversarial security challenges can be marked, establishing authority contexts, innovation with CyberSec concerns, funding of projects to institutions, and raising public awareness. The citation patterns interpret the intellectual structure of the field, pondering upon the early works leading to contemporary research and emerging paradigms.

### 1.2 Our contributions

We present a comprehensive scientometric analysis of AI-Steganography with SDG goals alignment. The research questions explored in this context include the following:

RQ1: What are the publication and citation trends within AI-steganography and explain the growth, peak, and decline in impact?

RQ2: What are the thematic clusters of AI-steganography, their evolution trends, and methodological approaches leading each cluster?

RQ3: Why only 18 out of 654 articles (2.75%) align with Sustainable Development Goals? What does this focused engagement tell us about the field's priorities, and what opportunities exist for broadening AI-steganography's contribution to societal challenges?

In this study, Section 2 outlines the structured methodology employed for data gathering and evaluation. Section 3 presents the results and discussion, followed by the conclusion in Section 4.

## 2. Research methodology

Scientometric analysis refers to the use of multiple qualitative and quantitative methods for academic sources analysis (Zarczynska, 2012). Researchers apply it to literary sources like journals, conference reviews, and books, among others, to identify emerging trends, future research scope, source analytics, and the intellectual structure of a specific domain (Donthu et al., 2021). We can analyse a vast and rich amount of scientific data to identify potential areas for future exploration. The availability of various scientometric analysis tools like VOSViewer (van Eck & Waltman, 2010), bibliometrics (Aria & Cuccurullo, 2017), scientific data sources like Scopus, Web of Science, Dimensions, Google Scholar, and data analytics tools like MS Excel and Power BI helps us to conduct bibliometric analysis at an articulate level. It squeezes the analysis process and focuses on the key trends, significant authors, and prime research clusters. Combining scientometric analysis with a statistical modelling tool for systematic literature review helps us summarize findings from available literature in a large dataset, with minimal loss of data quality and quantity. It increases efficiency by reducing noise and ensuring a comprehensive understanding of the research field.

This systematic literature review uses the Preferred Reporting Items for Systematic Reviews and Meta-Analysis (PRISMA) (Page et al., 2021). PRISMA has a checklist of 27 items that transparently explains and elaborates on the findings. Compared to other methodologies, the ease and clarity of implementing PRISMA help us answer our research goals lucidly. Figure 1 illustrates the study’s research design.

## 2.1 Search identification

In the context of this research, several databases were evaluated, like Web of Science, Scopus, Dimensions, IEEE, ACM Digital Library, and Google Scholar. However, Web of Science Core Collection (by Clarivate) and Scopus (by Elsevier) were identified for the scientometric study based on the quality of indexed papers, completeness of essential meta-data, and citation score (Singh et al., 2021). Also, including multiple databases might lead to analyzing a large data corpus with poor key referential information and irrelevant publications, affecting the quality of the study. We decided to use the words "Artificial Intelligence" in conjunction with "Steganography", along with other closely related technologies and synonymous keywords such as "Deep Learning", "Digital Watermarking", and "Concealed Data Transmission" to substantiate our study. Using only the primary technological terms limited our understanding of the field in broader and quantitative terms (about 25%) hence the additional terms were necessary for having a comprehensive and nuanced understanding of the subject matter. Our approach was limited to English sources because the analysis method used relies on Abstract data which might hamper the uniformity of the review method used. We specialized our study to Articles and Reviews considering their stringent features. Books, book chapters, and conference papers are excluded due to their commonly lower scientific impact and lower robustness (Kumar et al. 2023). It ensured the reliability and validity of the sources used, focusing on original research, and thereby maintaining the study's integrity. The source period from 2017-2023 is identified based on the publication trend. A total of 1070 publications, 697 from Scopus (675 articles & 22 reviews) and 373 publications from WOS (362 articles & 11 reviews) are identified.

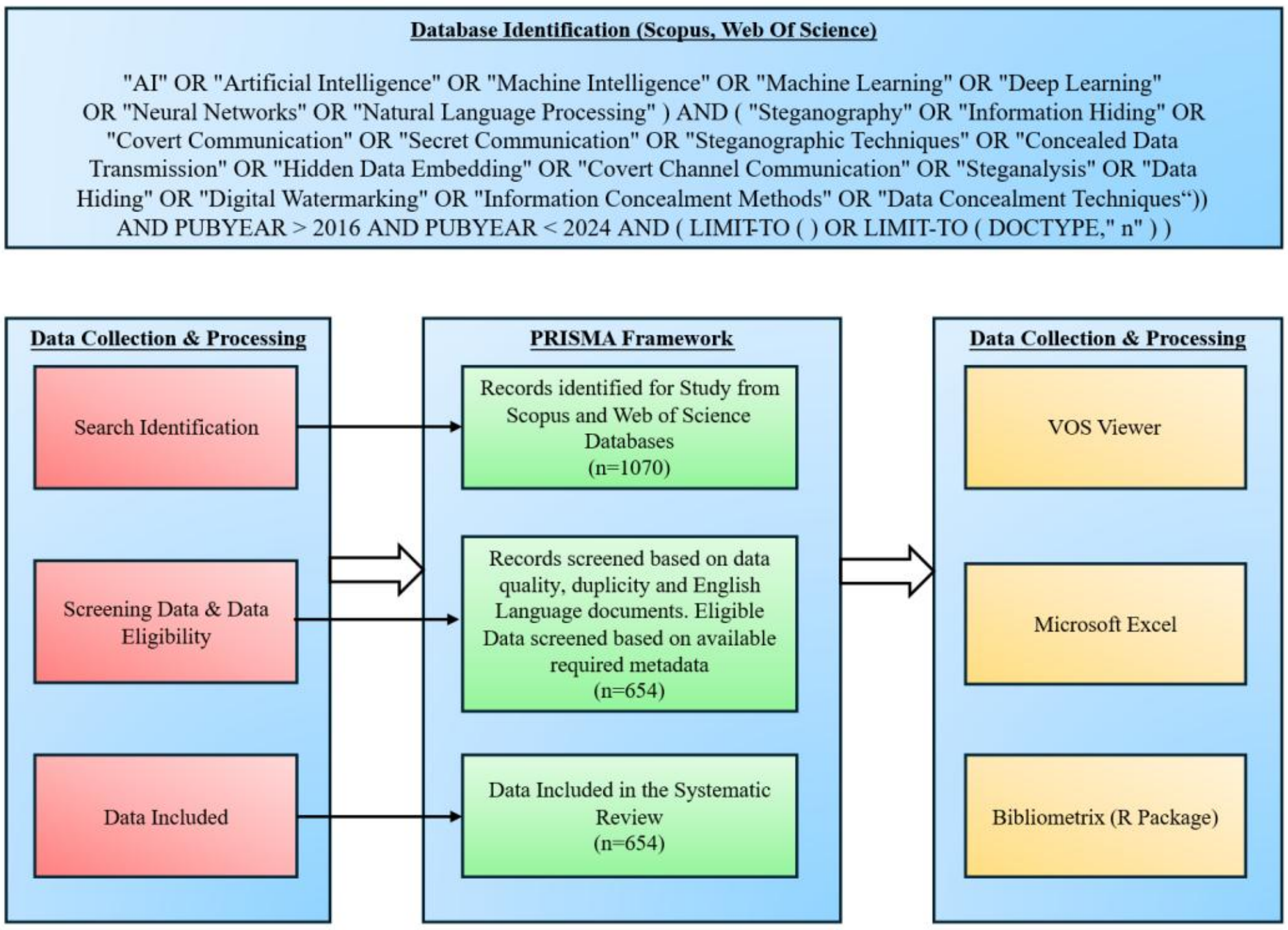

**Figure 1.** Research design of the proposed study

### 2.2 Screening data and data eligibility

Utilizing multiple data sources helps us get a holistic view of the domain for a scientometric analysis study, but it also comes with a challenge. Combining data sources might lead to duplication based on indexing by independent sources, defying our qualitative empirical. We utilized an R Package 'RefManageR'(W. McLean, 2017) to combine our search results from Web of Science and Scopus. It also permits us to merge metadata from both sources, fill in the missing data from either of the latter sources, remove duplicates, and combine different data file types to output in a single file, in a format of our choice. A total of 416 (duplicate or overlapped records of WoS & Scopus, a language other than English, Erratum, Conference, Conference review, retracted papers) records were removed on merging the records, leaving 658 unique records (~61.5 % data). We exported our data in a comma-separated value (.csv) file, which can be used for cross-platform analysis on Microsoft Excel, VOSViewer (van Eck & Waltman, 2010), and Bibliometrics (Aria & Cuccurullo, 2017) (tools used for analysis). The data was then manually analyzed in tabular format using Microsoft Excel to remove records missing pivotal metadata. 4 Erroneous/Incomplete Records were removed, leaving 654 records. An Eligibility Analysis was done based on the source publisher's reliability. Being a quantitative study as well, not many records were trimmed due to their usefulness in conducting bibliometric analysis and answering our research questions. In summary, 654 out of 1070 identified records (~61% unique, error-free data) were finalized after screening and data eligibility for this study.

## 3. Results and Discussion

### 3.1 Research Performance (RQ1)

A structured process was employed to fulfill the first research question of conducting a descriptive analysis of research performance in the field of AI-steganography. The selected data includes details on total publications, citations, leading authors, countries, affiliations, and top-cited papers. It also provides information on the top journals and sources associated with AI-steganography. By examining the publication trends, major milestones, and the evolution of AI-steganography can be tracked. The bibliometric indices offer a summarized look of the individual papers' impact, as well as the collective influence of authors on the field.

Microsoft Excel (MS Excel) was used for descriptive analysis. Bibliometrix, an R Package designed for bibliometric analysis, was used to extract relevant bibliometric metrics from the dataset for a numerical dissertation of the data. MS Excel, with Bibliometrix, was used for comprehensive data exploration, extracting essential perspectives on the key factors, articles, and the overall impact of AI-steganography research. Table 1 summarizes the data facts extracted from the tools.

**Table 1:** Summarized major facts

| | |
|---|---|
| Timespan | 2017:2023 |
| Sources (Journals) | 240 |
| Documents | 654 |
| Annual Growth Rate % | 65.3 |
| Document Average Age | 2.65 |
| Average citations per doc | 12.12 |
| References | 23010 |
| **Document contents** | |
| Keywords Plus (ID) | 3032 |
| Author's Keywords (DE) | 1517 |
| **Authors** | |
| Authors | 1466 |
| Authors of single-authored docs | 21 |
| **Authors collaboration** | |
| Single-authored docs | 27 |
| Co-Authors per Doc | 3.87 |
| International co-authorships % | 1.07 |
| **Document types** | |
| Article | 640 |
| Review | 14 |
| **Research Performance** | |
| G-index | 183 |
| H-index | 106 |
| i10-index | 181 |
| i20-index | 92 |

### 3.1.1 Annual Scientific Production and Citations

As part of RQ1, this section explains the article production and citation trends. The scientific production of publications has seen an annual growth of 65.3%, with an amplified momentum in 2018-2019 period. (Figure 2)

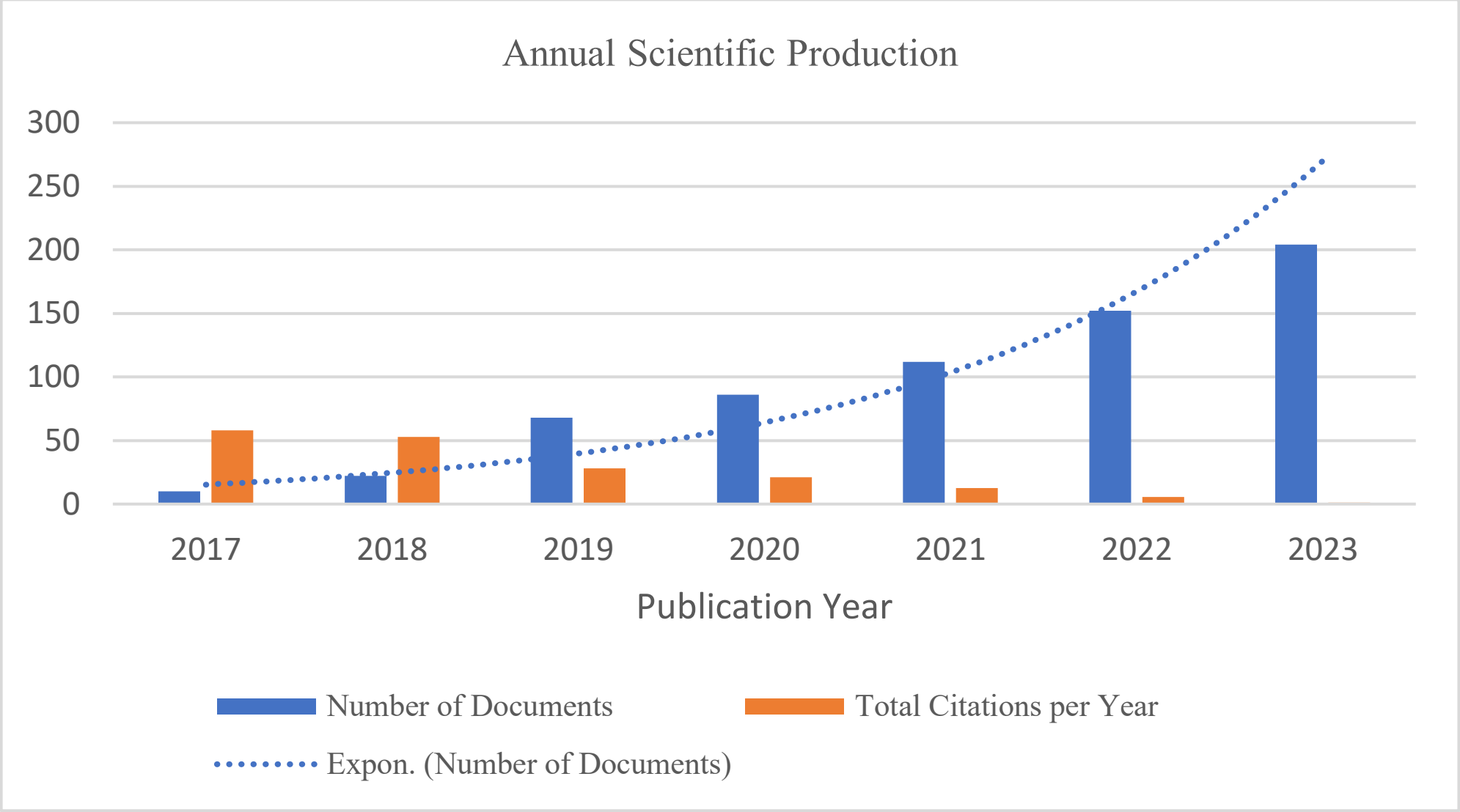

**Figure 2.** Annual growth of publications

**Table 2.** Publications and citations data

| Publication year | Total citations | Annual scientific production | Mean citations per year |
|---|---|---|---|
| 2017 | 578 | 10 | 57.8 |
| 2018 | 1162 | 22 | 52.8 |
| 2019 | 1908 | 68 | 28.1 |
| 2020 | 1807 | 86 | 21.0 |
| 2021 | 1390 | 112 | 12.4 |
| 2022 | 870 | 152 | 5.7 |
| 2023 | 214 | 204 | 1.0 |

#### 3.1.1.1 Interpreting the Publication and Citation Trajectory

There's a jagged rise in the number of works in 2018-2019 (22 to 68 articles per annum) (Fig. 2), which coincides with the mainstream embracing of Deep Learning frameworks and the explosion of Generative Adversarial Networks (GANs) in information hiding. These works fundamentally renovated steganography from handmade features to representations learned by models. The period signifies the rise of cyberattacks, in view of increased reliance on networking technologies, major data breaches, and investment in AI research.

The annual mean citations declined from 57.8 (2017) to 1.0 (2023), despite an increase in publication volume. It emphasizes the fact that the research landscape is mature, though fragmented, and dynamic. The early works formed the basis of the field and created the benchmarks, attracting persistent citations. In contrast, the explosion of publications in recent years and more methodological changes with the help of LLM tools reflect minor, though incremental improvements and niche applications, that are yet to gain significant traction and impact. The trend also points towards market saturation in the core areas (Eg, Image Steganalysis using Neural Networks), hence advising researchers to focus on emerging domains like Video Steganography, Linguistic Steganography, and Adversarial Robustness, which would take time to mature and gain recognition.

### 3.1.2 Research publications performance using H, G, I-10, I-20 index

The research performance analysis using researcher metrics (H, G, I-10, and I-20 indices) from 2017 - 2023 is presented in Figure 3 and Table 3. The peak of the G-index (43 in 2019) and H-index (23 in 2020), followed by sharp declines to 9 and 6, respectively, by 2023, shows the field establishment phase. The 2017-2020 period is advocated as the pinnacle of AI steganography research, where foundational deep learning methods (e.g., SRNet, CNN-based steganalysis, GAN-based embedding-less-steganography) were introduced and rapidly adopted across the industry and generating high citation velocity. The I-10 index (publications with >=10 citations) and I-20 index (publications with >=20 citations) further back these annual publication movements. The I-10 index for AI Steganography saw a sharp rise from 8 to 52 in the period 2017-2020, followed by a similar plummeting to 5 in 2023. The I-20 index culminated in 2020 at 29 from 3 in 2017, with a decline to 1 by 2023.

Papers published in 2021-2023 have had insufficient time to accumulate citations comparable to earlier works, especially given the 2-3 year citation window typical in computer science. AI Steganography has diversified into specialized subdomains (audio steganalysis, text steganography, quantum steganography), fragmenting the citation network and reducing the concentration of highly cited works. The decline in academic impact metrics may also reflect a transition from fundamental research to applied and commercial development, which generates fewer academic citations but greater practical utility.

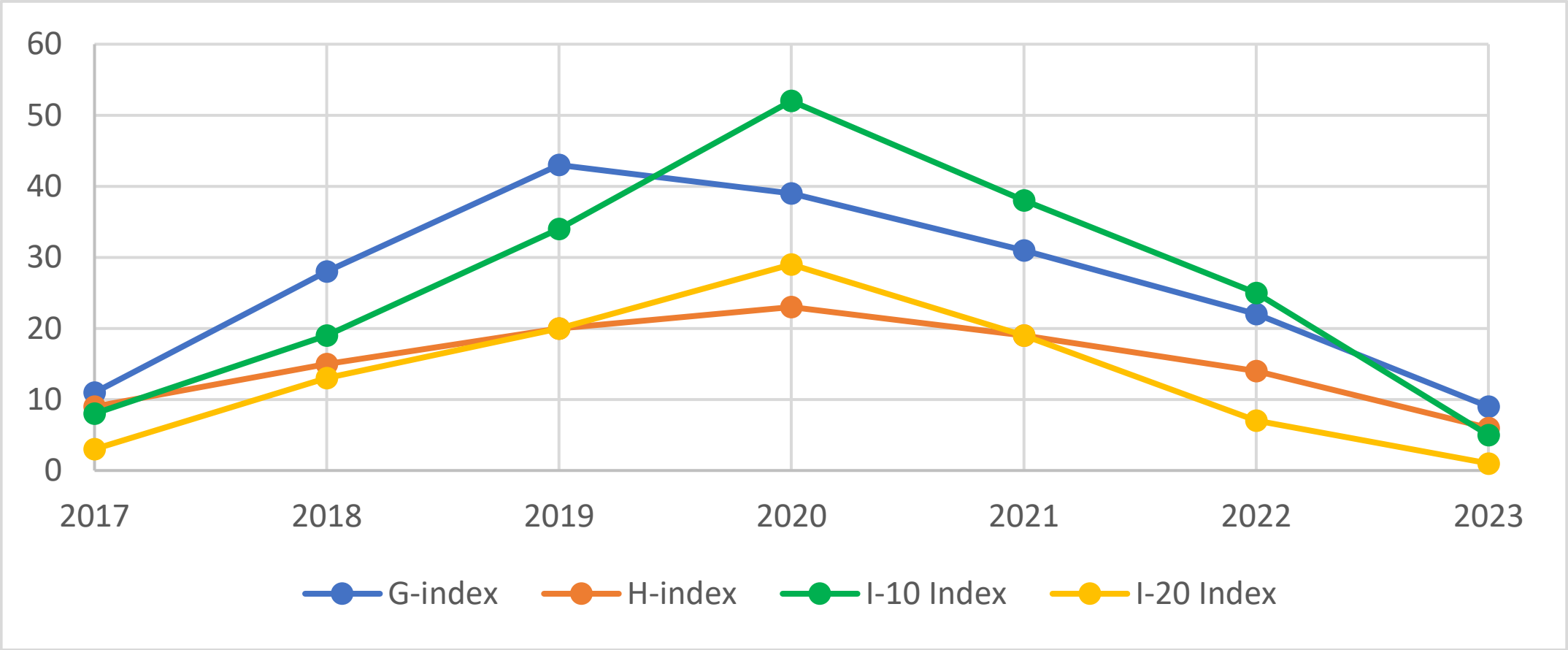

**Figure 3:** H, G, I-10, I-20 Index over the year

**Table 3:** H, G, I-10, I-20 Index

| Year | G-index | H-index | I-10 index | I-20 index |
|---|---|---|---|---|
| 2017 | 11 | 9 | 8 | 3 |
| 2018 | 28 | 15 | 19 | 13 |
| 2019 | 43 | 20 | 34 | 20 |
| 2020 | 39 | 23 | 52 | 29 |
| 2021 | 31 | 19 | 38 | 19 |
| 2022 | 22 | 14 | 25 | 7 |
| 2023 | 9 | 6 | 5 | 1 |

### 3.1.3 Top cited publications

Counting the number of citations is one way to assess the quality of a publication (Anderson 2006). Researchers have long used citation counts as a key metric to gauge the scientific impact and significance of research article publications (Bornmann and Daniel 2008). Table 4 summarises the top 10 cited documents of our corpus. The top two papers (Boroumand et al., 2019) and (Ye et al., 2017) have total citations (TC) of 461 and 455, with 92.2 and 65 citations per year, respectively. According to our observations, publications from 2017-2019 received a higher TC annually, indicating their significant impact. This is evident as the citation counts of older papers tend to be higher than those of newly published ones.

The top five papers based on citation counts focused on image steganalysis. Boroumand et al., (2019) talk about a novel deep residual network called SRNet for image steganography detection which aims to outperform existing methods by minimizing hand-designed elements and maximizing stego-signal preservation. It delivers excellent detection accuracy for JPEG steganography, usable in a wider range of applications. The architectural design of the proposed network comprises four layers, two involving the so-called residual shortcuts for improved convergence and parameter-based learning in the upper layers of deep networks. It has 3 linearly linked segments, the front for effective learning of 'noise residuals', the middle segment compacting the feature maps, and the final segment of a simple linear classifier. Ye et al., (2017) portray an alternative approach to the steganalysis of digital images based on CNN. The author's perspective, based on obtaining a complete description of the cover source via high-dimensional representation, is vital for making state-of-the-art steganalysis models. S. Wu et al., (2018) propose a new idea – a CNN model of deep residual network (DRN) for image steganalysis. The Deep Residual Network (DRN) has a large depth, which gives it a strong ability to understand the statistical properties of input data. Instead of learning a function directly, DRN approximates a residual mapping, helping to keep the weak signal produced by message embedding. Hu et al. (2018) devised a novel steganography without Embedding (SWE) method, which overcomes ML-based algorithm detection for image steganography. It is based on deep convolutional GANs (Generative Adversarial Networks), portraying a robust ability to tackle steganalysis detection and highly accurate information extraction. Tang et al., (2019) talk about three major works, a contemporary approach to counter ML-based steganalysis which isn't based on the attempt to safeguard a specific image statistical model, a practical steganographic scheme called ADV-EMB which can generate camouflage images capable of carrying secret information by adversarial embedding operation and an analysis of different adversarial models to achieve the optimal security metrics.

**Table 4.** Top 10 cited documents

| TC | Authors Name & Year | TC/Year |
|---|---|---|
| 461 | Boroumand et al., (2019) | 92.2 |
| 455 | Ye et al., (2017) | 65 |
| 247 | Shen et al., (2018) | 41.17 |
| 191 | Hu et al., (2018) | 31.83 |
| 175 | Yang et al., (2019) | 35 |
| 172 | Tang et al., (2019) | 34.4 |
| 157 | Zeng et al., (2018) | 26.17 |
| 148 | Zhang et al., (2020) | 37 |
| 136 | Yang et al., (2020) | 34 |
| 118 | Zhang et al., (2019) | 23.6 |

### 3.1.4 Top publishing journals

We found a total of 262 journals as publishers of all 654 scholarly works. The top 10 journals are shown in Table 5. Multimedia Tools and Applications journal by Springer tops our study with 58 articles/reviews, IEEE Access with 49 publications closely follows it, and IEEE Transactions on Information Forensics and Security (IEEE TIFS) with 22 publications. IEEE Access is the fastest growing source, showing a significant increase in 2018-2023, implying their focus on AI-steganography in recent years. Given the number of sources and production trends, many publications are spread across a wide variety of verified sources, indicating a growing trend and potential support from the research community. IEEE TIFS consistently ranks at the higher end of IF among the top venues, whereas IEEE Access and Multimedia Tools and Applications exhibit moderate IFs but high publication throughput and visibility.

**Table 5:** List of top journals

| TP | Journal Name | Impact Factor(IF) |
|---|---|---|
| 58 | "Multimedia Tools and Applications" | 3.0 |
| 49 | "IEEE Access" | 3.4 |
| 22 | "IEEE TIFS" | 8.0 |
| 21 | "IEEE Signal Processing Letters" | 3.9 |
| 18 | "Security and Communication Networks" | 1.29 |
| 17 | "Electronics (Switzerland)" | 2.6 |
| 12 | "IEEE Transactions on Dependable and Secure Computing" | 7.5 |
| 11 | "Journal of Information Security and Applications" | 3.7 |
| 11 | "Journal of Visual Communication and Image Representation" | 3.1 |
| 10 | "Journal of Electronic Imaging" | 1.0 |

### 3.1.5 Top Authors (by publication count)

Top authors published more than 15 articles is presented in Figure 4. We can notice striking contrast between the research productivity and long-term influence among the top authors. While Zhang Xinpeng is the most prolific author with 21 articles, Wang Xingyuan stands out as the most influential author, with a remarkably high H-index of 90. This indicates the weightage and impact of Wang's work having sustained impact on the concerned fields, with his papers receiving high number of citations over time.

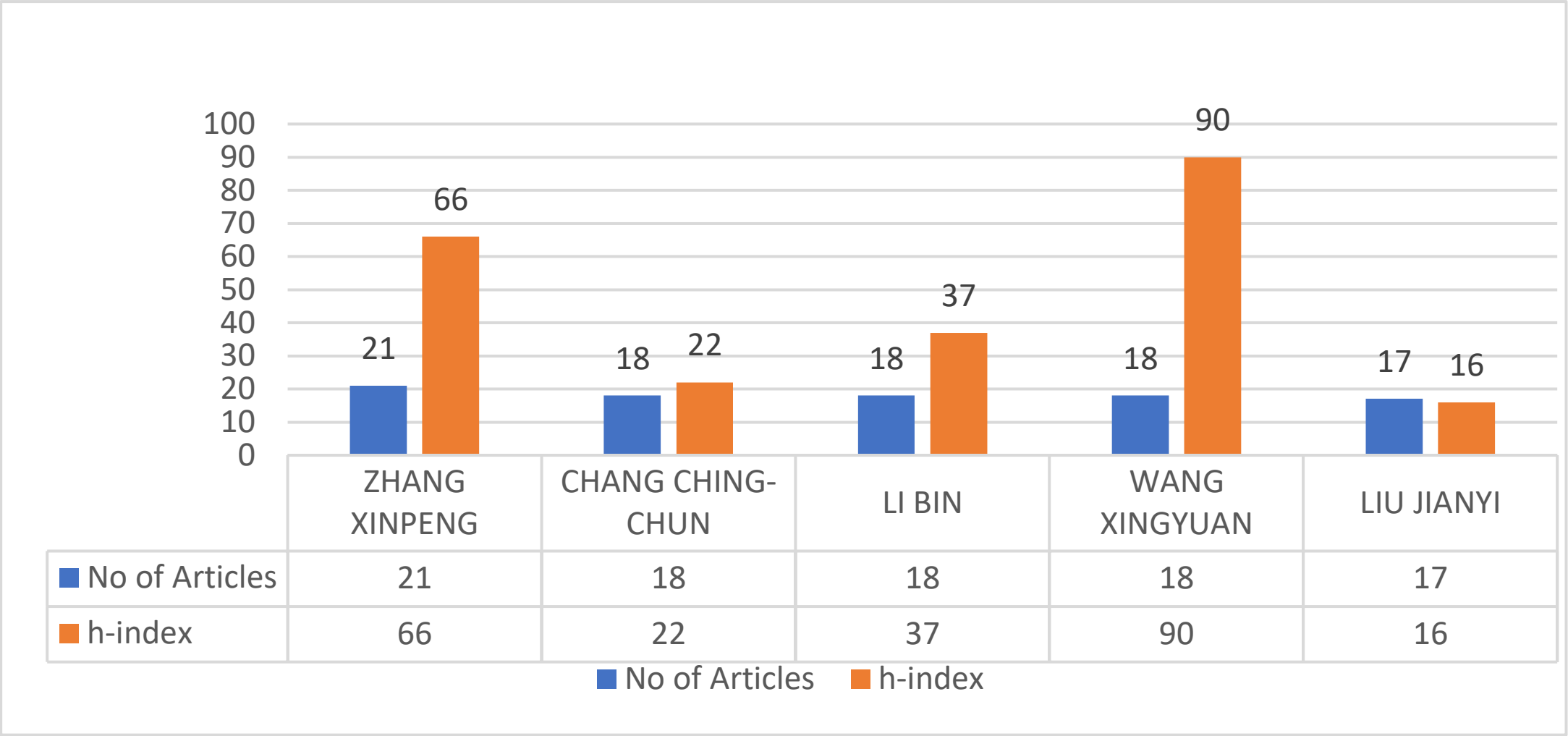

| | ZHANG XINPENG | CHANG CHING-CHUN | LI BIN | WANG XINGYUAN | LIU JIANYI |
|---|---|---|---|---|---|
| No of Articles | 21 | 18 | 18 | 18 | 17 |
| h-index | 66 | 22 | 37 | 90 | 16 |

**Figure 4:** Top Authors

It also reveals that having a similar article count doesn't guarantee a similar impact. For instance, authors Chang, Li B and Wang X all have 18 articles in AI Steganography field, but their h-indices vary from 22 to 90. This underlines the importance of using multiple metrics, to assess an author's contribution in the scientific community. Li B, with an H-index of 37, demonstrates a higher influence than Chang C, while Wang X's work is clearly the most influential amongst the AI Steganography landscape.

### 3.1.6 Top Institutions/Author Affiliations (by publication count)

Top institution by publication count is shown in Figure 5. It is observed that, Shenzhen University (China) leads in the terms of production with 67 articles. Followed closely are Sun Yat-Sen University (47 articles), Shanghai University (46 articles), and Nanjing University of Information Science and Technology (42 articles).

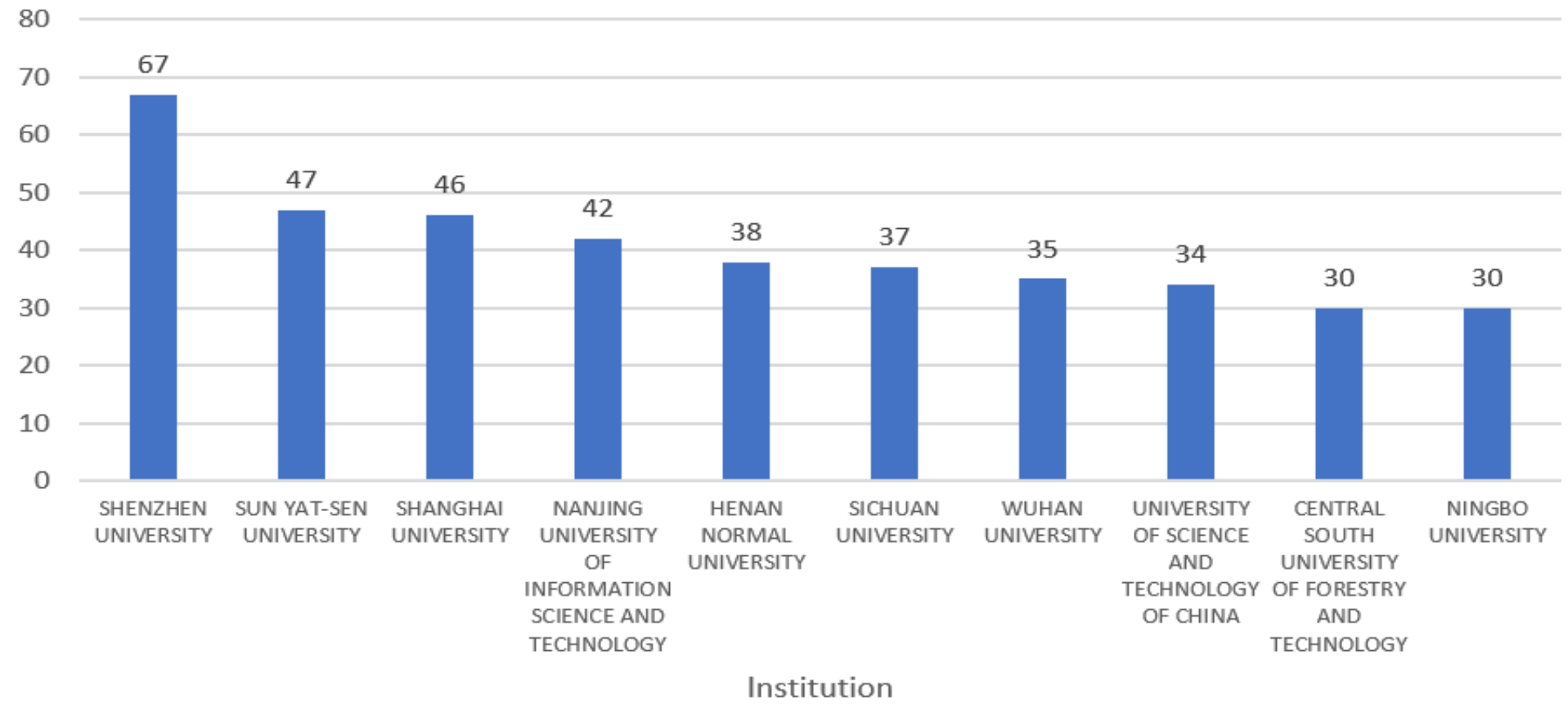

**Figure 5:** Top Institutions

The chart also shows a consistent level of output from several top institutions in China. Institutions like Nanjing University of Information Science and Technology and the University of Science and Technology of China are known for their technical excellence. This specialization indicates that research is being conducted by skilled teams of researchers and scientists within departments that have the necessary infrastructure and academic focus to excel in a niche field like AI steganography.

### 3.1.7 Top countries by scientific production and article citations

By Scientific Production, China, India, and the USA are on top with 312, 114, and 25 publications (Table 6). The 3/5th of the total production in this field implies an inclination of Asian countries. Total citations wise also China, India, and the USA are on top. We also notice less participation from African, North Asian, and European countries in the said field.

The geographic concentration in Asia, especially China (48% of total publications), manifests the strategic national investment in CyberSec and AI infrastructure seen in recent years, whilst the comparatively modest output from Europe and North America might signal different research priorities or a shift toward commercialization and deployment rather than academic publication.

**Table 6:** Top countries with Total Publications (TP) and Total Citations (TC)

| TP | TC | TC/TP | Country |
|---|---|---|---|
| 312 | 5080 | 16.3 | CHINA |
| 114 | 678 | 5.9 | INDIA |
| 25 | 568 | 22.7 | USA |
| 19 | 86 | 4.52 | AUSTRALIA |
| 12 | 75 | 6.3 | IRAN |
| 12 | 58 | 4.8 | EGYPT |
| 10 | 114 | 11.4 | JAPAN |
| 8 | 120 | 15.0 | SPAIN |
| 4 | 57 | 14.3 | FRANCE |
| 1 | 93 | 93.0 | QATAR |

### 3.1.8 Collaboration between countries

The countries' collaboration shed light on the country-specific efforts in the field of AI Steganography, with China emerging as the central and most prolific collaboration hub. Table 7 presents the country collaboration frequency, while Figure 6 shows the country collaboration map. China's partnerships are extensive, including major research nations, as the United States, the United Kingdom, and Australia. Outside China, the network stresses that Australia is a crucial liaison for multiple countries, like China, India, and the United Kingdom. The findings suggest that Australia is a significant Center of Excellence in AI Steganography research, which also rouses inter-country collaboration efforts across the Asia-Pacific and European Regions. As of now, we see a highly localised effort, with a handful of dominant, impactful players stressing on the country's goals and funding constraints as well.

**Table 7**: Country Collaboration Frequency

| From | To | Frequency |
|---|---|---|
| CHINA | AUSTRALIA | 134.4910001 |
| CHINA | IRELAND | -8.137935687 |
| CHINA | ITALY | 12.07001339 |
| CHINA | UNITED KINGDOM | -2.865631641 |
| CHINA | USA | -112.4616737 |
| INDIA | AUSTRALIA | 134.4910001 |
| UNITED KINGDOM | AUSTRALIA | 134.4910001 |
| UNITED KINGDOM | ITALY | 12.07001339 |

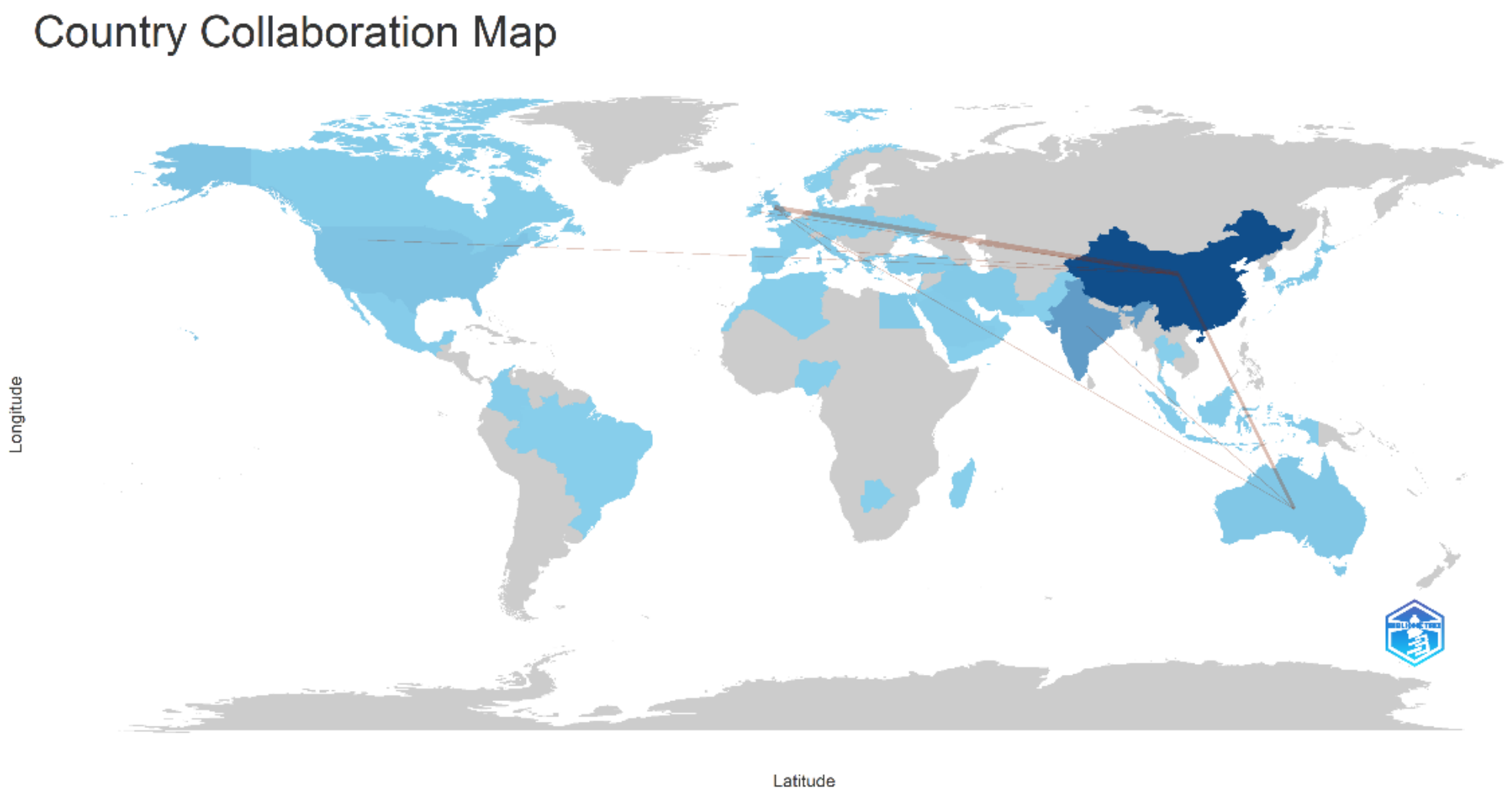

**Figure 6:** Country Collaboration Map

### 3.1.9 Top Funding Bodies

Figure 7 shows the top funding organisations in this domain of research. The National Natural Science Foundation of China overwhelmingly dominates, with 195 publications (~1/3rd of our selected data corpus).

**Figure 7:** Top Funding Organisations

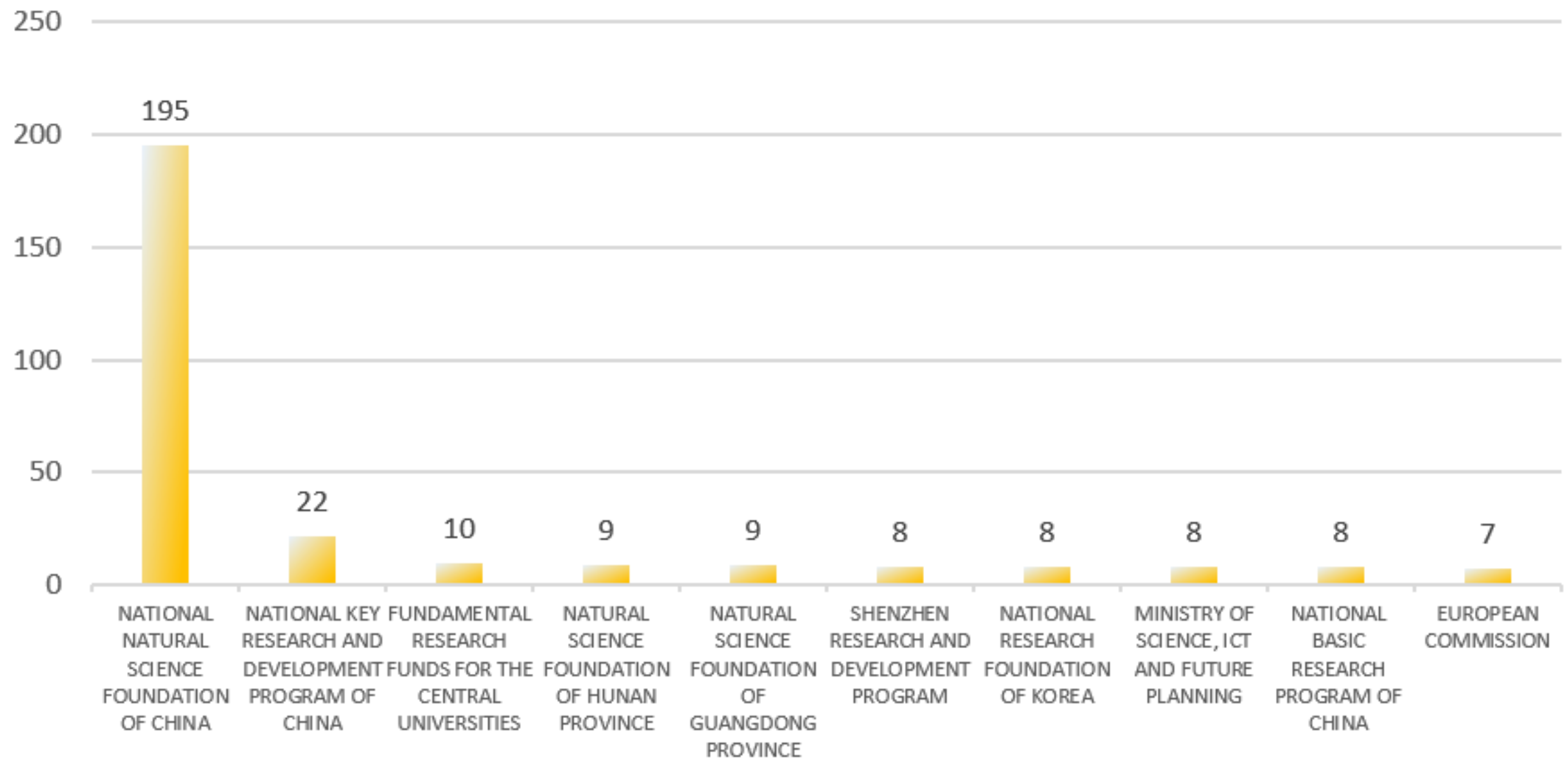

It is a solid multiple against the National Key Research and Development Program of China, with only 22 publications, as the second. The chart also shows that eight out of the top ten funding organizations are from China, with organizations, regional foundations, and universities providing immense support to the AI Steganography research. Apart from the strong Chinese presence, the data shows limited international representation, with the National Research and Development Foundation of Korea and the European Commission in the list. This data, along with our previous discussions on countries' collaboration and top countries, summarises that though the research is globally conducted, the funding for the publications is profoundly concentrated in China (or broadly, East Asia), showing its prowess in dynamic research output within the domain.

### 3.2 Thematic analysis (RQ2)

Thematic analysis is an essential method for uncovering key themes, involving the process of identifying, analysing, organizing, interpreting, and recording patterns found during data collection (Nowell et al. 2017). RQ2 directs on grouping research papers with similar abstracts. It identifies common themes and trends across various research areas, thereby unravelling a comprehensive, statistical understanding of the research field. By categorizing these papers into distinct clusters, we can comprehend the time evolution of the field, the interrelationships between research topics, and the progression of ideas and methodologies within the field. The objective not only aids in reviewing a vast amount of literature but also helps identify gaps in the present study, laying the groundwork for future studies. The VOS Viewer software tool for thematic cluster analysis constructs and visualizes bibliographic networks (Van Eck and Waltman, 2014). Seven clusters have been identified based on the abstract with a minimum of 10 occurrences. 440 terms were identified, out of which the top 60% (264 words) with a total link strength of 71275 were used for cluster analysis (Figure 8).

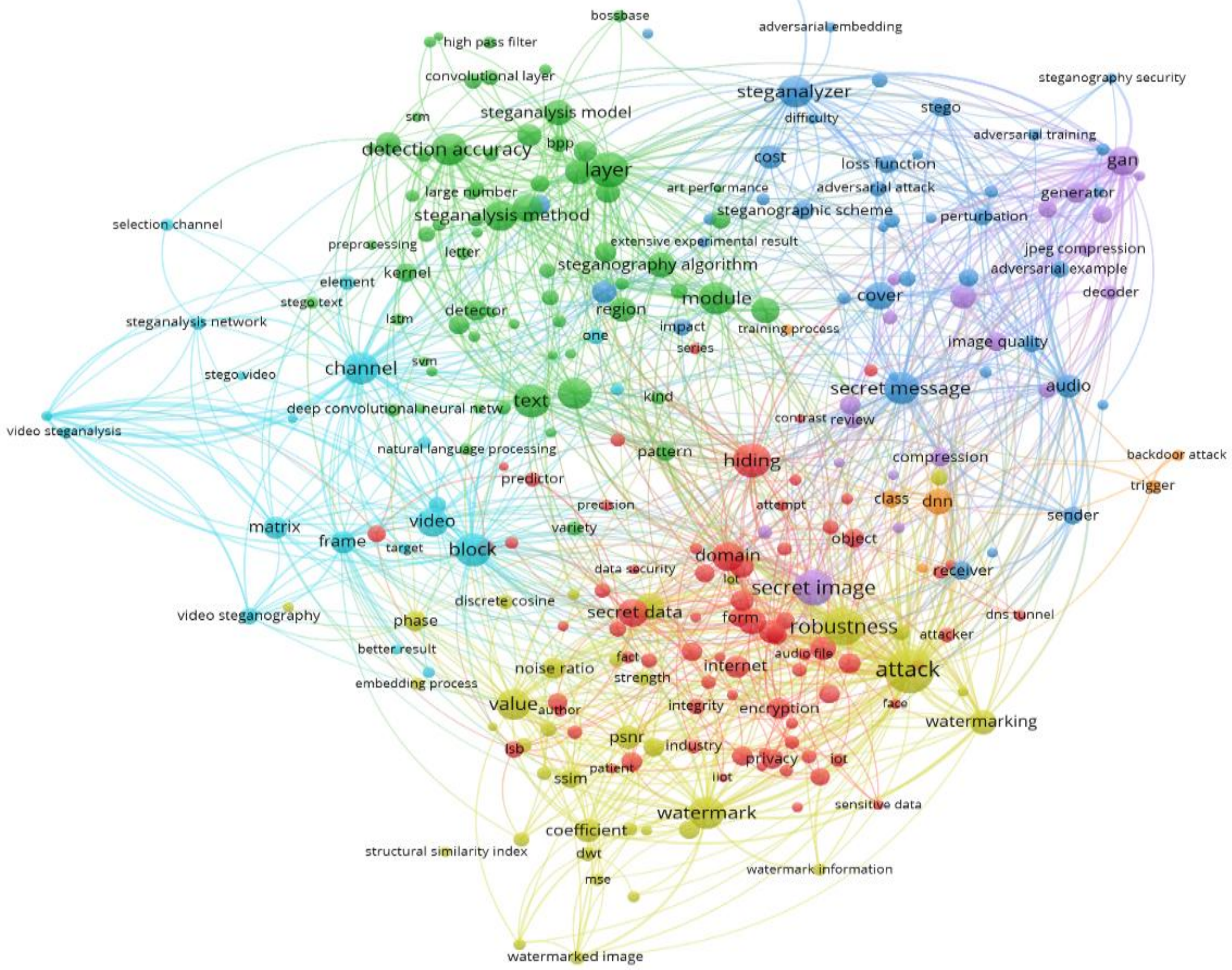

**Figure 8:** Thematic Clusters

***Cluster 1 (Red): Encryption-based data hiding***

Cluster 1 is dominated by 'hiding', 'domain', 'secret data', 'encryption', 'cryptography', and 'environment', suggesting that the data points to data security, encryption, and possibly environmental factors affecting these processes.

With the explosive amplification of data, there is a challenge in storing as well as sharing the data. Apart from that, the integrity of the data is also of supreme importance. Concerning this, Shen et al. (2018) suggested an identity-based shared remote data integrity approach that focuses on achieving data integrity in cloud storage. In general, the identity-based shared remote data integrity approach obtains a signature for each data block before storing it in the cloud. Further, the authors have used a sanitizer for sanitizing each data block along with signatures. Abd El-Latif et al., (2018) suggested two new quantum data hiding-based approaches. In this, the carrier quantum image hides another quantum image. At first, the secret quantum image is enciphered using a controlled-NOT gate and the two most significant and least significant qubits. Also, using Arnold's cat map, the secret quantum image is scrambled to preserve anonymity. The obtained image offers acceptable visual as well as perceptual quality while maintaining justifiable hiding capacity. Research in steganography-based data hiding is principally centred around maintaining a fair trade-off among the steganography measures, such as robustness and hiding capacity. In this aspect, a stego key-directed two-level encryption algorithm in combination with the adaptive least significant bit (LSB) is suggested by Muhammad et al. (2017). The proposed work can be divided into four different phases, such as (1) encrypting the secret key using the two-level encryption approach, (2) Next, encrypting the secret information using the multi-level encryption algorithm, (3) Embedding the enciphered secret data in different channels of the image using the adaptive LSB approach, and (4) extracting the encrypted embedded information using the exact opposite occurrences.

The rise of encryption-based approaches in this period was mainly due to the advancing need for addressing trust deficits in cloud and IoT environments, where data integrity and confidentiality are crucial. The rise of quantum computing threats also sparked interest in quantum steganography as a future-proof answer to unbreakable security measures. The cluster's sustained presence across the timeline stresses its role as a basic building block, upon which today's advanced techniques are developed. Finally, the experimental results reveal that the given approach achieves a rational balance among the conflicting measures, such as high capacity, perceptual quality, and security.

***Cluster 2(Sky): Video Steganography Techniques:***

Cluster 2 has the highest frequency of 'channel', 'video', 'block', 'video steganography', and 'matrix', suggesting fields of video processing and video steganography.

Pirated versions of a legal copy are significant issues for content creators as well as the entertainment industries. Piracy is a copyright infringement, and it can be combated with technological solutions. Liu et al. (2019) suggested a comprehensive review and analysis of digital video steganography techniques for secure video transmission. In this, the authors have classified the video steganography techniques into three different classifications, such as (1) intra-embedding (IE), pre-embedding (PE), and post-embedding (POE). The IE technique considers intra-prediction, estimation of motion in a video, and discrete cosine transform coefficients for embedding the secret data in a video. The PE video steganography can be implemented in both spatial as well as transform domains. In the case of POE video steganography, the message is concealed in the compressed video bitstream. Protection of digital video using watermarking is one of the rapidly growing fields of research. In this regard, to combat the piracy issue in the case of both 2D and 3D videos, Asikuzzaman and Pickering (2018) performed an in-depth analysis of its application and challenges. Generally, the 3D watermarking methods are partitioned into three different categories: 3D/3D, 3D/2D, and 2D/2D. Further, the 3 videos can be represented as (i) stereo imaging, (ii) depth image-based rendering, and (iii) multi-view imaging systems. Mstafa et al. (2017) proposed a robust video steganography technique using multiple object tracking and error debugging codes in both discrete wavelets transform and discrete cosine transform domains. At first, the secret information is pre-processed. Next, using the multiple objects tracking algorithm, the secret information is embedded in the region of interest of the video. Since the information is concealed in a moving object, the frames of a video usually keep changing over every instance of time leading to provide robust steganographic approach. Further, a remarkable breakthrough in the domain of quantum data hiding is proposed by Naseri et al. (2017). The suggested work utilizes the quantum comparator circuit to hide confidential information in the least significant as well as most significant bits using the XORing technique. The results of the suggested technique reveal that the suggested work improves both hiding capacity as well as superior perceptual quality.

The shift from image to video-based hiding reflects the practical need for higher embedding capacity and the challenge of evading steganalysis in temporally dynamic content. Anti-piracy measures and intellectual property protection in the entertainment industry further drove this trend. However, the computational complexity of processing video streams and the limited availability of standardized datasets have constrained rapid growth compared to image-based methods.

***Cluster 3(Yellow): Robust Watermarking Techniques***

Cluster 3 has 'robustness', 'attack', 'watermarking', 'phase', 'psnr', 'watermark', and 'imperceptibility'. This suggests a relation to robustness in data security, encryption, and watermarking techniques.

Information hiding secures and maintains the confidentiality of electronic patient information. In this respect, Anand and Singh (2020) suggested an improved watermarking technique using the DWT- Singular value decomposition (SVD) domain to transmit the patient record over public channels. The DWT is used to partition an image into different sub-bands, namely (1) approximation (LL), horizontal (LH), vertical (HL), and diagonal (HH). The authors suggested that the LL sub-band region is more of a smoother region and hence it is not favorable for hiding any information. On the other hand, the HH sub-band can resist comparably additional information. The carrier image is encrypted and then compressed using several approaches during embedding. The authors have pointed out that the Hyper Chaotic-Lempel-Ziv-Welch (LZW) has performed superior attack resistance ability against different subjective and objective attacks. Shehab et al. (2018) devised a block authentication-based self-recovery approach for enhancing medical image authentication. Initially, the covered watermark image is partitioned into 4×4 blocks for embedding, and the authentication bits are obtained using the SVD and LSB substitution approach. The suggested SVD-based fragile watermarking not only detects the tampered regions from the watermarked image but also recovers the altered region with excellent precision. In the recent wake of the COVID-19 pandemic, the medical care industry has embraced remote patient diagnosis, monitoring devices, and electronic patient records for quicker and robust health protection. In this connection, for an efficient telemedicine system ensuring confidential transmission a 3D medical-based watermarking system is suggested by Zhang et al. (2022a). The proposed work utilizes a combination of principal component analysis, particle swarm optimization, and bacterial foraging model for achieving optimal trade-offs among the conflicting measures.

The emphasis on robustness against attacks and imperceptibility aligns with regulatory requirements for patient data protection (HIPAA, GDPR). The alignment of several papers in this cluster with SDG 3 (Good Health and Well-being) and SDG 9 (Industry, Innovation, and Infrastructure) underscores the cluster's societal relevance, though it remains underexplored relative to the field's total output.

***Cluster 4 (Blue): Audio Steganalysis Techniques***

Cluster 4 has major frequency words like 'Steganalyzer', 'Audio', 'Secret Message', 'Trigger Sender', 'Perturbation', 'Cost', and 'Cover' implying Steganalysis, Audio Processing, Secret messaging, Trigger Mechanisms, Perturbations, Cost Factors, and Cover Mechanisms.

Digital health platforms, such as telemedicine, telehealth, and online pharmacy services have been rapidly adopted after the COVID-19 pandemic. However, infringement in any of the digital platforms can pose significant risks to the privacy of the patient leading to no longer having trust in the overall electronic healthcare system. In this context, to protect the medical audio information, Zhang et al. (2022b) suggested a two-stage reversible secure audio watermarking approach. The medical audio information is embedded in two different domains, respectively. In this, the authors ensured that even with the attack on the medical audio information, the embedded information could still be retrieved without any tampering. This ensures that the given work holds significant resistance to different watermarking attacks. Hwang et al. (2017) advocated for an improved SVD-based audio watermarking technique that utilizes the quantization index modulation to protect the original audio signal. The authors demonstrated that the produced singular values are connected with the audio signal to achieve better imperceptibility as well as security. Further, the suggested technique successfully withstands the volumetric scaling attack. The data transmission in audio files is susceptible to noise. A high compression-based audio watermarking has been suggested by Ali et al. (2018) The suggested approach utilizes the combination of fractal coding and chaotic LSB-based hiding. Here the inceptive measures for the chaotic map are considered as the private keys for data communication. The computational evaluation for the proposed work suggests that the obtained output image has superior transparency and fidelity as compared to its counterparts.

Audio steganalysis emerged as a critical area around 2019-2021 driven by the proliferation of voice-based interfaces (smart assistants, VoIP) and concerns about covert channels in IoT devices. The rise of this cluster reflects defensive research aimed at detecting malicious data exfiltration in audio streams. However, it remains a smaller cluster due to the relative maturity of image and video steganography, which attract more research attention and funding.

***Cluster 5(Violet): Adversarial Image Generation***

Cluster 5 has words like 'GAN', 'Secret Image', 'Discriminator', 'Decoder', 'Generator', and 'Encoder' implying a focus on fields like Generative Adversarial Networks (having parts Encoder and Decoder), Secret Image (a concept of Steganography).

Conventional data-hiding techniques usually leave traces of the inclusion of concealed information in an image. Therefore, these techniques show modest robustness against modern machine learning (ML) based detection tools. In this regard, Hu et al. (2018) suggested an embedding-less steganography strategy using deep convolutional generative adversarial networks to resist contemporary ML and AI based steganalysis techniques. The suggested approach consists of 3 phases. At first, the deep convolutional generative adversarial networks are trained to obtain the generator. The generator produces the cover image. Then the model is trained again with an extractor to confirm the synchronization between the generator and extractor networks. The devised mode has been trained with both the Celebrities and Food101 datasets of 200k face images and 50k food images, respectively. The results from different forensic experiments reveal the strong anti-steganalysis ability of the suggested model. Further, to fool the CNN-based steganalyzer, a novel adversarial embedding approach to improve the embedding capacity has been suggested by Tang et al. (2019). Yang et al. (2029) proposed a distortion function, generating a steganography framework. It consists of three modules: a generator, a no-pretraining-required double-tanh function, and an enhanced CNN-based steganalyzer, together with multiple high-pass filters as the discriminator. The proposed UT-6HPF-GAN framework outperforms the state-of-the-art steganography methods, including HILL, MiPOD, S-UNIWARD, and ASDL-GAN on other datasets. The key innovation—steganography without embedding (SWE)—addresses the fundamental weakness of traditional methods that leave statistical traces. This cluster has had an outsized impact (many highly cited papers fall here) because it fundamentally redefines the steganography-steganalysis arms race.

***Cluster 6(Green): Text Detection Module***

Cluster 6 focuses on 'Layer', 'Module', 'Detection Accuracy', 'Convolutional Layer', 'Classifier', and 'Text', related to topics of text detection using ML and DL.

The text detection module in steganography serves as a critical component for identifying and extracting hidden textual data embedded within digital media, ensuring that covert communication remains intact and undetected by conventional analysis tools. Elhoseny et al. (2018) proposed a hybrid security model for safeguarding the diagnostic text data embedded in medical images. The hybrid encryption scheme developed in this work combines both the Advanced Encryption Standard (AES) and Rivest-Shamir-Adleman (RSA) algorithms. The model encrypts the confidential data, which is then embedded into a cover image utilizing either 2D-DWT-1L or 2D-DWT-2L techniques. The method accommodates both color/grayscale images as cover mediums to obscure varying sizes of text. The system's effectiveness was assessed using six statistical metrics: peak signal-to-noise ratio (PSNR), mean square error (MSE), bit error rate (BER), structural similarity index (SSIM), structural content (SC), and correlation. Yang et al. (2018) proposed an approach for linguistic steganography i.e., RNN-Stega. This method utilizes Recurrent Neural Networks(RNN) to automatically produce high-quality text covers based on a secret bitstream that requires embedding. Fixated-Length Coding (FLC) and Variable-Length Coding (VLC) are used for encoding words based on their conditional probability distribution. A substantial amount of artificially created samples was used to train the model, and the performed experiments show the proposed model surpasses all prior related approaches and attains advanced performance. The transmission of patient records via a network necessitated a method to ensure the security and confidentiality of tele-health services. Anand & Singh (2020) introduced an enhanced watermarking method that can safeguard patient data by incorporating multiple watermarks into medical cover images using the DWT-SVD domain. To mitigate channel noise distortion for sensitive data, the Hamming code is applied to the text watermark before embedding.

This cluster remains in the "emerging or declining" quadrant of the thematic map, indicating limited traction. This may be due to the inherent difficulty of generating natural, semantically coherent steganographic text and the relatively low embedding capacity compared to multimedia methods.

***Cluster 7 (Orange): Deep Neural Networks (DNN) based Backdoor Attack Detection***

Cluster 7 has the words 'Trigger', 'DNN', 'Class', New Method', and 'Backdoor Attack', suggesting Deep Neural Networks (DNN), classification methods, new methodologies, and possible security concerns related to backdoor attacks.

Backdoor attacks, where malicious actors embed hidden vulnerabilities in our system that cause specific misbehaviour, pose significant threats to our systems. ML and Big Data Analytics (BDA) serve as powerful tools for analyzing and securing our systems. Gyamfi & Jurcut (2022) support the argument of how ML-BDA can benefit us in analyzing and securing technology. They conducted a literature review of existing ML-based detection solutions and detailed a case study with a real-world sandbox for conducting cyber-attacks and designing an intrusion detection system (IDS). Nadler et al. (2019) introduced a method for detecting DNS-based malware, emphasizing low-throughput techniques that had been largely overlooked in previous studies. The authors proposed a solution that handles streaming DNS traffic for detecting and denying requests to domains used for data exchange. The method is designed to collect DNS logs over long periods and extract features based on each domain's querying behavior for malware detection, and develop an anomaly detection model to classify domains according to their use for data exfiltration. Kumar et al. (2021) proposed a Fog computing-based, distributed ensemble IDS to protect IoT networks. It tells how IoT has developed alongside Fog computing, and an IDS is necessary to detect modern IoT attacks. The proposed IDS uses a distributed architecture, combining KNN, XGBoost, and Gaussian Naive Bayes at the first level, and Random Forest at the second level for final classification.

The 2020-2023 works align with the scrutiny emphasis of responsible AI and trustworthiness, especially in security-first applications. The major focus areas are intrusion detection systems (IDS), malware detection, and DNS-based exfiltration indicates a defensive stance, as AI steganography is used in malicious intents. The cluster's limited size suggests it is still in an early developmental stage.

**Table 8:** Cluster Analysis with Influential Methods

| Cluster | Influential Methods | Analysis |
|---|---|---|
| Encryption-based data hiding (Cluster 1) | Identity-based shared remote data integrity approach , Quantum data hiding (controlled-NOT gate, Arnold's cat map) , Stego key-directed two-level encryption with adaptive LSB | Sustained presence till date, since upward trend from 2017 onwards |
| Video Steganography Techniques (Cluster 2) | Intra-embedding (IE), Pre-embedding (PE), Post-embedding (POE) , Multiple object tracking/ECC in DWT-DCT domains | A trend shift from image to video has been observed in recent years (2021 onwards) |
| Robust Watermarking Techniques (Cluster 3) | DWT-SVD domain watermarking , Hyper Chaotic-Lempel-Ziv-Welch (LZW) , SVD-based fragile watermarking , PCA/PSO/Bacterial Foraging 3D medical watermarking | Aligns with rise of telemedicine/patient data protection regulations (2019 onwards) |
| Audio Steganalysis Techniques (Cluster 4) | Two-stage reversible secure audio watermarking , SVD-based adaptive QIM watermarking , Fractal coding/chaotic LSB-based hiding | 2019–2021 (Proliferation of voice-based interfaces and IoT covert channels) |
| Adversarial Image Generation (Cluster 5) | Steganography without Embedding (SWE) using deep convolutional GANs , Adversarial embedding , UT-6HPF-GAN framework | 2018–2019 (Coincides with mainstream embracing of Deep Learning and GANs) |
| Text Detection Module (Cluster 6) | Hybrid encryption (AES/RSA) with 2D-DWT-1L/2L , RNN-Stega (Recurrent Neural Networks for text cover generation) | Limited Traction, thereby open to trends of emergence or decline in the coming years |
| Deep Neural Networks (DNN) based Backdoor Attack Detection (Cluster 7) | ML-BDA , DNS-based malware detection , Distributed ensemble IDS (KNN, XGBoost, GNB, RF) | 2020–2023 (Scrutiny of responsible AI/trustworthiness) |

### 3.2.1 Thematic Map

The thematic map is used to visualize the relationship between themes based on their relevance and developments over time (Tennekes, M., 2018). A thematic map is a very intuitive plot, and we can analyse themes according to the quadrant in which they are placed: 1) upper-right quadrant: motor themes; 2) lower-right quadrant: basic themes; 3) lower-left quadrant: emerging or disappearing themes; 4) upper-left quadrant: very specialized/niche themes.

***Motor themes:*** Themes in the upper-right quadrant are known as the motor themes, characterized by high centrality and high density, meaning that they are developed and important for the research field (Cobo et al., 2022; Aria et al., 2020). From Figure 9, we can observe that cryptography, digital watermarking, stego characterized steganography using deep learning are coming under motor themes.

***Basic and transversal themes:*** Themes in the lower-right quadrant are known as basic and transversal themes, characterized by high centrality and low density, meaning that these themes are important for a domain, and they concern general topics transversal to the different research areas of the field (Cobo et al., 2022; Aria et al., 2020). Figure 9 depicts that, image enhancement, image steganography, and NN are under basic themes.

***Emerging or declining themes:*** Themes in the lower-left quadrant are known as emerging or declining themes, with low centrality and low density, meaning that are not fully developed or marginally interesting for the domain (Cobo et al., 2022; Aria et al., 2020). Figure 9 shows that, natural language processing systems, linguistics, and text steganography fall under Emerging or declining themes.

***Niche Theme:*** Themes in the upper-left quadrant are known as the high developed and isolated themes, with well-developed internal links (high density) but unimportant external links (low centrality), meaning that strongly developed but still marginal for the domain under investigation (Cobo et al., 2022; Aria et al., 2020). Figure 9 highlights that, algorithms, NNs, and computer algorithms are under niche themes.

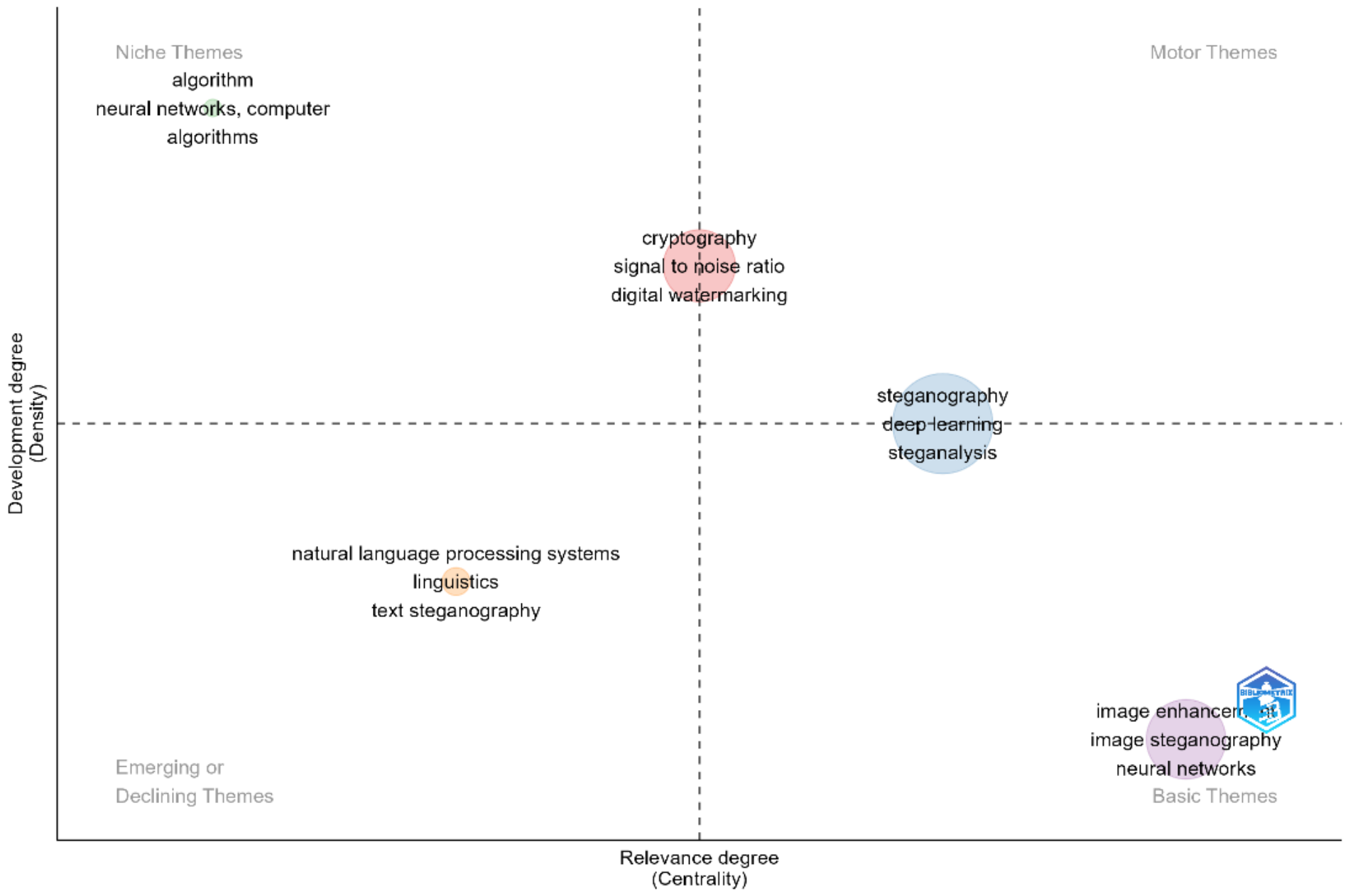

**Figure 9.** Thematic Map

### 3.2.2 Thematic Evolution

Figure 10 highlights the thematic evolution of AI-steganography. The period of 2017-2019 focuses on foundational technologies like "cloud computing", "information hiding", and "neural networks", making the beginning of AI-steganography. In 2020-2021, the focus includes "deep learning", "linguistic steganography", "digital watermarking", "reversible data hiding", and "motion vector", indicating a shift towards more advances and specialized areas of research, with increased emphasis on security and privacy. The predicted themes for 2022-2023, include "steganography", "reinforcement learning", "security", "intellectual property protection", and "multi-task learning", suggesting a continued focus on security and privacy, with increased emphasis on AI and ML techniques.

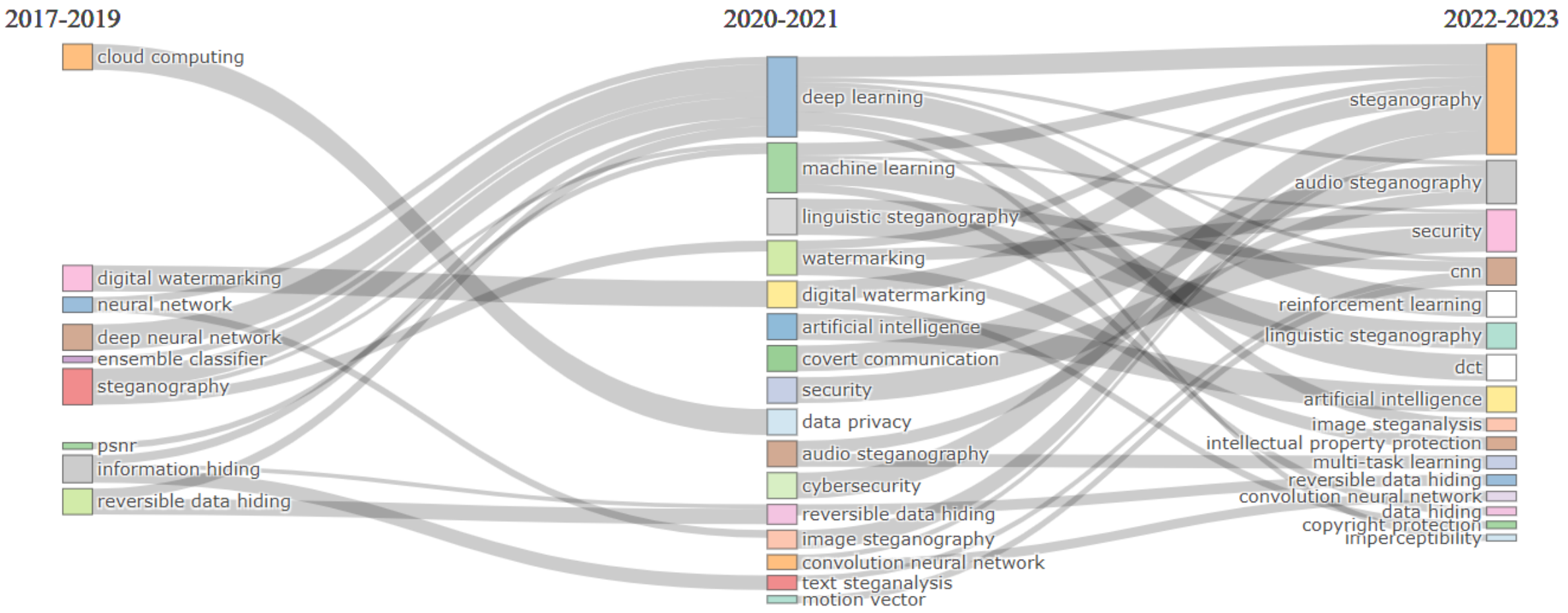

**Figure 10.** Thematic Evolution

### 3.3 SDG Mapping (RQ3)

The last objective of our research is to investigate the correlation between articles and Sustainable Development Goals (SDGs) through SDG mapping (Raman et al. 2024a). Scopus and SciVal are mostly utilized to correlate papers with Sustainable Development Goals (SDGs). The SDGs were established by the United Nations in 2015. The goals encompass a total of 17 objectives that are required to be accomplished by the year 2030 (Raman et al. 2024b). The goal progression is assessed using 169 indicators (Raman et al. 2024c).

Only 18 of the 654 AI-steganography articles clearly align with Sustainable Development Goals (SDGs), primarily SDG 9 (Industry, Innovation, and Infrastructure).

This scarce alignment is because maximum research prioritizes technical advancements—such as 'enhancing embedding capacity' and 'detection accuracy', over 'societal impact'. The field's focus is on core security and encryption challenges, with the traditional academic incentives. Therefore, it calls for interdisciplinary engagement to pressing SDG issues like health, inequality, or climate action. Furthermore, many applications contribute wholly to SDGs, without explicit acknowledgment, which causes underreporting. In summation, this research gap shows missed opportunities for AI steganography, addressing paramount global challenges, and calls for awareness, interdisciplinary collaboration, and stronger incentives to frame research work within sustainable development contexts.

**Table 9:** Identification of SDG mapping

| Authors & Year | TC | SDG |
|---|---|---|
| Hassaballah et al., (2021) | 82 | SDG 9 |
| Jeong & Park (2019) | 38 | SDG 11 |
| Jain et al., (2021) | 35 | SDG 9 |
| Cai et al., (2022) | 26 | SDG 7 |
| Wang et al., (2023) | 22 | SDG 9 |
| Zhang et ai., (2022) | 9 | SDG 10 |
| Sahu et al., (2020) | 7 | SDG 7 |
| Yazdanpanah et al., (2023) | 7 | SDG 3 |
| Acer et al., (2022) | 4 | SDG 9; SDG 11; SDG 16 |
| Hoda & Mondal (2022) | 3 | SDG 16 |
| Alissa et al., (2023) | 3 | SDG 9 |
| Dhawan & Gupta (2021) | 3 | SDG 16 |
| Ragab et al., (2022) | 2 | SDG 9 |
| Amrit & Singh (2023) | 2 | SDG 11 |
| Narayana et al., (2022) | 2 | SDG 9 |
| Zhang & Zhang (2023). | 2 | SDG 9 |
| Alghazzawi et al., (2022) | 1 | SDG 3 |

## 4. Conclusion

A comprehensive scientometrics analysis has been conducted within the field of AI-steganography to know the publications and citation trends (RQ1). We performed the thematic analysis using keyword co-occurrence to examine AI-driven steganography-based data hiding methods, each offering unique methodological advantages. Thematic analysis provided a clear, structured approach for identifying key themes, ensuring that the main research areas were well-defined and comprehensible (RQ2). The study mainly identifies seven clusters, namely (1) encryption-based data hiding, (2) video steganography techniques, (3) robust watermarking techniques, (4) audio steganalysis techniques, and (5) adversarial image generation, (6) text detection module, (7) deep neural networks (DNN) based backdoor attack detection. Our SDGs mapping (RQ3) analysis identified 18 articles linked to the Sustainable Development Goals (SDGs). The majority of highly cited and impactful articles align with SDG9, which focuses on industry, innovation, and infrastructure.

**Limitation of the study:**

Our study faces certain limitations. First, the chosen database may not capture the full spectrum of research on the topic, especially the work published in regional journals, smaller conferences or non-English publications. Second, while the use of thematic modelling offers a strong analytical framework, each method has its own limitations in capturing the full complexity of the field, possibly leading to the oversimplification of some elements. Additionally, the interpretation of themes, even with advanced modelling, can be subjective, introducing potential biases or misinterpretations, which highlights the importance of expert review and validation. Lastly, as this study relies on citation counts related to identified clusters to gauge impact, it is subject to the well-known shortcomings of citation analysis.

**CRediT authorship contribution statement:** Conceptualization, Methodology, Investigation: Aditya Kumar Sahu, Chandan Kumar. Writing – original draft: Aditya Kumar Sahu, Chandan Kumar, and Saksham Kumar. Writing – review & editing: Aditya Kumar Sahu, Chandan Kumar, Serdar Solak. Data curation & Software: Chandan Kumar and Saksham Kumar. Supervision: Chandan Kumar

**Declaration of competing interest:** The authors declare no known competing personal interests that would influence the work reported in this paper.

**Acknowledgments:** The research is supported by the AMRITA Seed Grant, Amrita Vishwa Vidyapeetham, Amaravati campus (File Number: ASG2022234).